\documentclass[mathleft
]{an}
\usepackage[]{natbib}
\usepackage{graphicx}
\usepackage{times}
\usepackage[english]{babel}
\bibpunct{(}{)}{;}{a}{}{,}

\newcommand\aap{{A\&A}}%
\newcommand\aaps{{A\&AS}}%

\begin{document}

\Pagespan{141}{}
\Yearpublication{2009}%
\Yearsubmission{2008}%
\Volume{330}%
\Issue{2/3}%
\DOI{10.1002/asna.200811141}%

\title{Parsec-scale Jet in the Distant Gigahertz-Peaked Spectrum Quasar PKS\,0858$-$279}

\author{Y.~Y.~Kovalev\inst{1,2}\fnmsep\thanks{Corresponding author:
  \email{ykovalev@mpifr-bonn.mpg.de}\newline}
}
\titlerunning{GPS quasar PKS\,0858$-$279}
\authorrunning{Y.~Y.~Kovalev}
\institute{Max-Planck-Institut f\"ur Radioastronomie,
Auf dem H\"ugel 69, 53121 Bonn, Germany
\and
Astro Space Center of Lebedev Physical Institute,
Profsoyuznaya 84/32, 117997 Moscow, Russia
}

\received{2008 Oct 30}
\accepted{2008 Dec 18}
\publonline{2009 Feb 15}


\abstract{%
The high redshift GPS quasar PKS\,0858$-$279 exhibits the following properties
which make the source unusual. Our RATAN-600 monitoring of 1--22~GHz
spectrum has detected broad-band radio variability with high amplitude
and relatively short time scale. In the same time, the milliarcsecond
scale structure observed in a snapshot VLBA survey turned out to be very
resolved which is not expected from the fast flux density variations. We
performed 1.4--22 GHz VLBA observations of this quasar in 2005--2007. 
It has revealed a core-jet morphology. A high Doppler factor $\delta$ is
suggested for the jet, its nature is discussed in this report on the
basis of the multi-frequency VLBA and RATAN data collected. Synchrotron
self-absorption was confirmed to be dominating at low frequencies,
the magnetic field strength of the dominating jet feature is estimated
of an order of $0.1\delta$~mG.
}

\maketitle

\section{Introduction}

At RATAN-600 we monitor more than 600 extragalactic objects North of
declination $-30^\circ$ since 1997 \citep{Kovalev_etal99}. We have
identified and analyzed a subsample of objects which permanently show
gigahertz-peaked spectra (GPS; see \citealt{S09}). One of the GPS
sources in the sample, high redshift quasar PKS\,0858$-$279 ($z=2.152$,
\citealt{SKF93}) was already identified as a GPS source and studied in
many papers including
\cite{SPG85,OBS91,CSCQ94,VBH95,VBD97,EdwardsTingay04}. This distant
object has drawn our attention because of several properties
which are unusual for both, a typical GPS source and a typical quasar.
First results of a study of this interesting quasar are presented in
this paper.

We adopt for this paper
the Friedmann-Robertson-Walker cosmology with the following parameters:
$H_0 = 72$~km\,s$^{-1}$ Mpc$^{-1}$, $\Omega_M = 0.3$, and $\Omega_\Lambda = 0.7$.
For a source at a redshift $z=2.152$, the luminosity distance
$d_\mathrm{L}=16.5$~Gpc, one milliarcsecond corresponds to 8~pc.

\section{Properties of the GPS quasar 0858$-$279}
\subsection{Variable radio spectrum}

\begin{figure}[t]
\centering
\includegraphics[trim=0cm 0cm 0cm 0cm,width=0.45\textwidth]{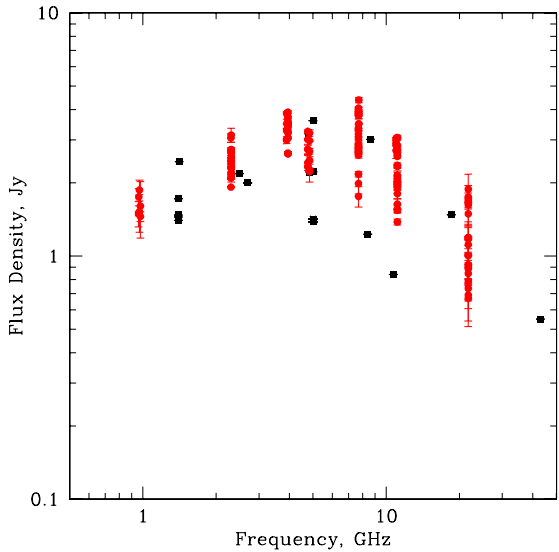}
\caption{\label{f:0858sp}
Variability of the broad-band spectrum of the GPS quasar 0858$-$279.
Filled circles are RATAN observations in 1997--2008 at frequencies
1.0, 2.3, 3.9, 4.8, 7.7, 11, \& 22 GHz, filled
squares come from the literature (collected by the CATS database).
RATAN light curves are presented in Figure~\ref{f:0858lc}.
}
\end{figure}

The quasar 0858$-$279 shows GPS-like spectral shape at all observing
epochs (Figure~\ref{f:0858sp}). Our analysis of variability properties
has selected this quasar as one of the most variable objects among the
identified subsample of GPS sources.
This finding is supported by
ATCA monitoring results \citep{EdwardsTingay04}. They have shown that
the variability index for this object is the highest one among the GPS
sample investigated (we do not consider 1519--273 which is known to be a
strong IDV and NGC\,1052 which is not a ``true'' GPS). Most of GPS
sources which possess a simple convex spectrum at any observing epoch
are not variable significantly.

We also note that this object is reported by \cite{Ricci_etal04} to
have ($106\pm7$)\,mJy of linearly polarized total flux density at
18.5\,GHz which makes its fractional polarization ($7.1\pm0.5$)~per~cent. Such high
degree of linear polarization is quite unusual for a flat spectrum radio
source and even more unusual for a GPS one.

\subsection{Characteristic milliarcsecond-scale size and morphology}

\begin{figure}[t]
  \centering
  \includegraphics[angle=270,trim=0cm 0cm 0cm 0cm,width=0.49\textwidth]{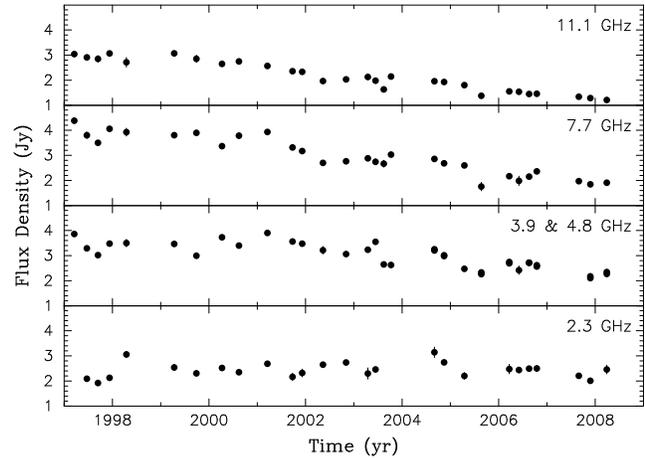}
\caption{\label{f:0858lc}%
Multi-frequency RATAN-600 light curves for the quasar 0858$-$279.
}
\end{figure}

We have used data from the VLBA Calibrator Survey program
\citep{VCS1,VCS2,VCS3,VCS4,VCS5,VCS6} to analyze the milliarcsecond
scale structure of the quasar. It did not show the expected core-jet
morphology and was found to be significantly resolved: the dominating
component is 1.3~mas at 8.6~GHz and 12~mas at 2.3~GHz in size (width at
half-power level of a Gaussian model component). In opposite, the shortest
time scale of the radio variability observed is of the order of several
months (Figure~\ref{f:0858lc}) which gives an estimation of the size of
the varying component from light-travel time arguments in the absence of
coherence \citep{Marscher_etal79} to be much smaller than the measured
sizes! The amplitude of the radio variability is about or greater than
1\,Jy. 

\section{Revealing the jet at parsec scales}

\begin{figure*}[p]
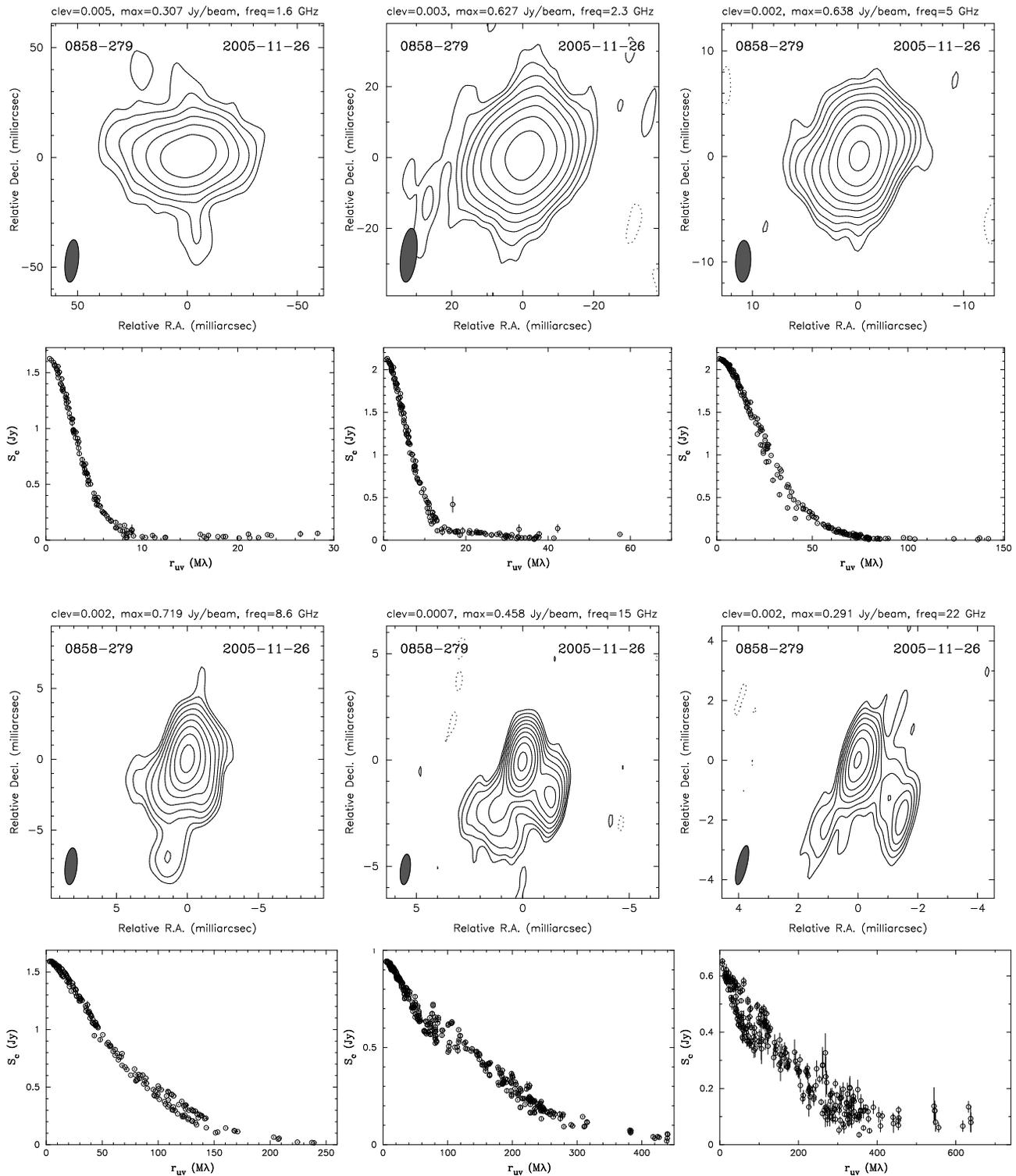

\centering
\resizebox{1.0\hsize}{!}{
  \includegraphics[angle=0,trim=0cm 5.5cm 0cm 0cm]{Kovalev_fig3a.eps}
  \includegraphics[angle=0,trim=0cm 5.5cm 0cm 0cm]{Kovalev_fig3b.eps}
  \includegraphics[angle=0,trim=0cm 5.5cm 0cm 0cm]{Kovalev_fig3c.eps}
}
\resizebox{1.0\hsize}{!}{
  \includegraphics[angle=270,trim=0.8cm    0cm 0cm 0cm,clip]{Kovalev_fig3g.eps}
  \includegraphics[angle=270,trim=0.8cm -0.3cm 0cm 0cm,clip]{Kovalev_fig3h.eps}
  \includegraphics[angle=270,trim=0.8cm -0.3cm 0cm 0cm,clip]{Kovalev_fig3i.eps}
}
\resizebox{1.0\hsize}{!}{
  \includegraphics[angle=0,trim=0cm 5.5cm 0cm 0cm]{Kovalev_fig3d.eps}
  \includegraphics[angle=0,trim=0cm 5.5cm 0cm 0cm]{Kovalev_fig3e.eps}
  \includegraphics[angle=0,trim=0cm 5.5cm 0cm 0cm]{Kovalev_fig3f.eps}
}
\resizebox{1.0\hsize}{!}{
  \includegraphics[angle=270,trim=0.8cm    0cm 0cm 0cm,clip]{Kovalev_fig3j.eps}
  \includegraphics[angle=270,trim=0.8cm -0.3cm 0cm 0cm,clip]{Kovalev_fig3k.eps}
  \includegraphics[angle=270,trim=0.8cm -0.3cm 0cm 0cm,clip]{Kovalev_fig3l.eps}
}
\caption{\label{f:0858VLBAres}
From top to bottom.
{\em First and third row:}
Naturally weighted CLEAN images of the quasar 0858$-$279 from 1.6 to 22
GHz. The lowest contour is plotted at the level given by ``clev'' in
each panel title (Jy/beam), the peak brightness is given by ``max''
(Jy/beam). The contour levels increase by factors of two. The dashed
contours indicate negative flux. The beam is shown in the bottom left
corner of the images.
{\em Second and fourth row:}
Dependence of the correlated flux density on projected
spacing from 1.6 to 22 GHz. Each point represents a coherent
average over one several minute observation on an individual interferometer
baseline. The error bars represent only the statistical errors.
}
\end{figure*}

\begin{figure}[t]
\centering
\includegraphics[angle=0,trim=0cm 0cm 0cm 0cm,width=0.49\textwidth]{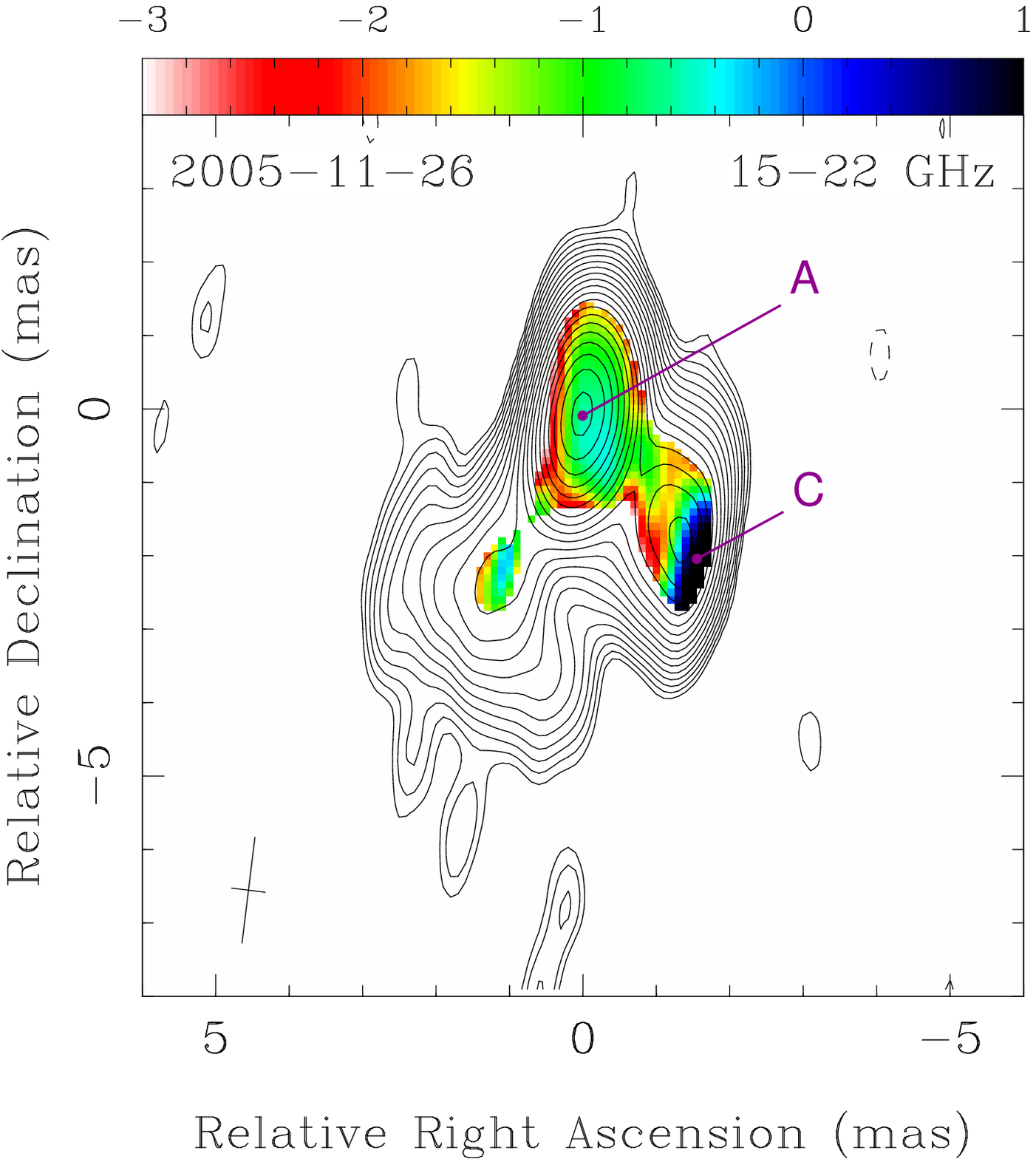}
\caption{\label{f:0858spmap}
Spectral index image of the quasar 0858$-$279 between 15 and
22~GHz combined with a CLEAN countor image made at 15 GHz.
Spectral index $\alpha$ is calculated assuming 
$S\propto\nu^\alpha$.
`C' marks position of the opaque core region while `A'~--- dominating
jet component which becomes optically thin above about 10~GHz.
}
\end{figure}

In order to resolve the obvious inconsistency between the variability
and milliarcsecond-scale structure, a dedicated VLBI experiment was
organized by us. We have observed the GPS quasar simultaneously in six
frequency bands (1.6, 2.3, 5, 8.6, 15, 22~GHz) with the NRAO VLBA array
in a dual circular polarization mode during four observing epochs in
2005--2007. First epoch results are presented in
Figure~\ref{f:0858VLBAres} (Stokes I), the spectral index image for the
highest frequency range observed~--- in Figure~\ref{f:0858spmap}. We
have found the following.

It is confirmed that the structure is heavily resolved at all observing
frequencies. It becomes more compact at higher frequencies while the 
dominating component of the emission becomes optically thin. The size of
the dominating feature A is measured from a $uv$-data analysis. It changes
from 26~mas at 1.6 GHz (optically thick case) to about 0.5~mas at 22~GHz
(optically thin case). This evolution of size with frequency is well
illustrated by the correlated flux density data presented in
Figure~\ref{f:0858VLBAres}. It is important to note that in this
particular case different VLBA resolution at different
frequencies did not significantly affect or limit accuracy of our size estimates.

The spectral index distribution calculated between 15 and 22~GHz shows a
compact component C with inverted spectrum ($\alpha\sim+1$,
$S\propto\nu^\alpha$) which is most probably an opaque core region while
the dominating component A with $\alpha\sim-1$ must be a jet region of the
quasar (Figure~\ref{f:0858spmap}). The high resolution high frequency
observations have finally revealed a typical core-jet structure at mas
scale for this quasar; the jet appears highly curved.

Electric vectors of linear polarization are found to be parallel to the
apparent inner jet direction at high radio frequencies and
rotate by about 90 degrees from 15 to 8-5~GHz. At the same time,
fractional linear polarization of A drops from 10 to 20~per~cent at high
frequencies to about or less than 4~per~cent at 8~GHz and below. This
suggests the synchrotron radiation with self-absorption from a blob with
highly ordered magnetic field. Magnetic field lines are therefore
perpendicular to the direction from C to A.

We have measured the true angular size of the dominating component at
frequencies above the synchrotron self-absorption turnover frequency
which appeared to be 0.5--0.6\,mas. The observed strong variability of
the total flux density with the variability time scale between two years
and 100 days requires a very high Doppler factor
$\delta>10$ \citep[calculated following][]{Marscher_etal79,Burbidge_etal74,Agudo_etal06}
and, from a probability argument \citep[e.g.,][]{Cohen_etal07},
a small viewing angle for this distant quasar.

VLBA monitoring of the structure of this quasar during about two years
in 2005-2007 did not allow us to detect any significant motion in the
structure. The distance between the core and the dominating jet
component remained about constant at 15 and 22~GHz.
Assuming a ``standard'' model for a compact synchrotron source
\citep[following the orignal idea by][]{Slysh63} we estimate the
magnetic field strength of the dominating jet component A having Doppler
factor $\delta$ \citep{M83,M87} to be of an order of $0.1\delta$ mG.

\section{Summary}

An unusual combination of properties of the high redshift GPS quasar
PKS\,0858$-$279, namely relatively fast flux density variations and 
``amorphous'' parsec-scale structure was resolved by VLBA observations
at high radio frequencies which revealed a parsec-scale core-jet
morphology for this AGN. The dominating jet component (A) responsible for the
GPS-type spectrum became optically thin above 15 GHz and was resolved
out from the core (C). From fast variations it is suggested that the jet
Doppler factor $\delta$ may be about or greater than 10. Linear
polarization analysis has confirmed that the turnover in the spectrum
of the dominating jet feature A is caused by the synchrotron
self-absorption, its magnetic field strength 
is estimated to be of an order of $0.1\delta$~mG.

The nature of the dominating jet component A with synchrotron
self-absorption still remains unclear. A feature with such
properties is unusual for the majority of the radio loud quasar jets.
What causes this change of jet properties, does interaction with the
surrounding medium play a role? How may short time scale variations
occur in the optically thick part of the spectrum in this quasar?
Further investigations will shed light on this.

Although formally this object can be classified as GPS source both based
on the spectral as well as on the linear-size definition, the new
observations reveal that this classification can probably no longer be
maintained based on the high boosting factor, core-jet morphology, and
other characteristics, which are unusual for a typical GPS source, but not
uncommon for a quasar.

\acknowledgements
We thank Margo and Hugh Aller for observing 0858$-$279 and calibrators
quasi-simultaneously with the VLBA experiment which helped to
calibrate polarization data.
The author is grateful to two anonymous
referees for their constructive comments which helped to improve the
manuscript.
The National Radio Astronomy Observatory is a facility of the National
Science Foundation operated under cooperative agreement by Associated
Universities, Inc.
This work is based in part on results from VLBA projects
BK\,128, BK\,133.
\mbox{RATAN-600} observations are partly
supported by the Russian Foundation for Basic Research (projects
01-02-16812, 05-02-17377, 08-02-00545).
Y.~Y.~Kovalev is a Research Fellow of the Alexander
von Humboldt Foundation.
The author made use of the
database CATS \citep{CATS05} of the Special Astrophysical Observatory.
This research has made use of NASA's Astrophysics Data System and the
NASA/IPAC Extragalactic Data\-base (NED).


\end{document}